\pdfoutput=1
\documentclass{article}

\usepackage{amsmath}
\usepackage{graphicx}
\usepackage{float}
\usepackage{hyperref}

\newcommand{\bolds}[1]{\mbox{\boldmath $#1$}} 
\title{Modeling Excess Deaths After a Natural Disaster with Application to Hurricane Maria}
\author{Roberto Rivera\thanks{College of Business, University of Puerto Rico, Mayaguez} \and  \and Wolfgang Rolke\thanks{Department of Mathematical Sciences,  University of Puerto Rico, Mayaguez}}
\begin{document}

\maketitle

\abstract{Estimation of excess deaths due to a natural disaster is an important public health problem. The CDC provides guidelines to fill death certificates to help determine the death toll of such events. But, even when followed by medical examiners, the guidelines can not guarantee a precise calculation of excess deaths.
We propose two models to estimate excess deaths due to an emergency. The first model is simple, permitting excess death estimation with little data through a profile likelihood method. The second model is more flexible, incorporating: temporal variation, covariates, and possible population displacement; while allowing inference on how the emergency's effect changes with time. The models are implemented to build confidence intervals estimating Hurricane Maria's death toll.
}	

\section{Introduction}

Estimating the death count due to an emergency, such as a natural disaster, is an important problem with far reaching consequences. For
example, in the United States, a death due to a natural disaster determines whether the family is eligible
for assistance from FEMA, the U.S. government disaster aid organization.
Also, the death toll influences the perspective of other countries of the
severity of the situation, and therefore their willingness to help. Most importantly, with the information residents can make educated, potentially life saving decisions. Many studies have assessed excess deaths due to some emergency.\cite{flegaletal05, nishiuraandchowell08, kutneretal09} Methods used include: descriptive summaries,\cite{brunkardetal08}
death notice models,\cite{stephensetal07} or log linear models.\cite{granetal13} The CDC provides guidelines on how to include disaster relatedness in death certificates.\cite{cdc18} However, arriving at an accurate total of indirect deaths by examination of each death is ultimately challenging. For example, a person dies a week after the natural disaster from a heart attack.
Did this heart attack happen because of storm aftermath stress? Or would it have happened
anyway? The answer is not always clear cut. Moreover, emergency conditions may quickly exacerbate noncommunicable diseases, increasing mortality risk.\cite{demaioetal13, ryanetal16}

On the morning of September 20, 2017, Hurricane Maria slammed into Puerto Rico with sustained winds of 155 mph and higher gusts. The effects of wind and rain from the cyclone were felt for over 24 hours. The atmospheric event led to devastation across the island; residents were left with no running water, no electricity, and no form
of communication. The severe infrastructure damage hindered proper forensic protocol to determine causes of death. Reportedly, after bodies started to pile up in the
morgues, hundreds were cremated without proper examination.\cite{prakash17}

The official government figure of Maria's death toll, which included direct and indirect deaths, was 64. But several reports concluded that the actual number was likely to be at least ten times higher.\cite{santosetal18, santosandhoward18, kishoreetal18, riveraandrolke18, roblesetal17,  sutteretal17} We apply two proposed models to compare the number of death certificates in Puerto Rico before and after the
storm, and determine how many more deaths occurred after the storm than one
would expect from historical mortality rates. 

\section{Method}

\subsection{Model 1: The Profile Likelihood Method}
A reasonable approach is as follows. Let $X_i =$ daily deaths in $m$ days before the emergency, and $Y_j =$ daily deaths in first $n$ days after the emergency; 
Each of these random variables are independent, and follow a Poisson distribution. Let's call $\lambda$ the ``background" rate for the daily deaths before the emergency. The emergency resulted in an additional ``source" of deaths, say at a rate $\rho$, for given post-Maria time periods. We can model the number of deaths after the emergency as following a $\text{Poisson}(\lambda + \rho)$ distribution. The goal is to estimate excess deaths as a function of $\rho$.
Say we have $X_{1},...X_{m} \sim \text{Poisson}(\lambda)$, and $Y_{1},...Y_{n} \sim \text{Poisson}(\lambda + \rho)$. The likelihoods will be a function of the sum of the random variables, which assuming independence, also follow a Poisson distribution. Set $x =\sum_{i=1}^{m}x_i$ and $y =\sum_{j=1}^{n}y_j$. The profile likelihood method is a useful way to perform estimation in the presence of nuisance parameters.\cite{coleetal13} In our case, defining the log-likelihood function as $l(\lambda,\rho;x,y)$, we set $\rho$ to a given value, $\rho_o$, and then find the value of $\lambda$ that maximizes the log-likelihood $l(\rho_o)= \sup_{\lambda}l(\lambda,\rho_o,;x,y)$. The computation is repeated across a broad range of values for $\rho$ and each time, we obtain the maximum of $l(\rho_o)$. For the likelihood ratio test, under the null $\rho=\rho_o$, it can be shown that (see Appendix):


\begin{equation*}
\widehat{\widehat{
		\lambda }} (\rho_{o})=\left( x+y -(m+n) \rho_{o} \pm \sqrt{(x+y-(m+n)\rho_{o})^{2}+4(m+n)x\rho_{o} }\right) /\left( 2(m+n)\right)
\end{equation*}

The profile likelihood ratio - the log of the profile likelihood divided by the likelihood evaluated at the maximum likelihood estimator - is:
\begin{equation}
\begin{split}
-2\log \text{LRT}(\rho_{o}) &= 
-2\left[ \log L(\widehat{\lambda},\widehat{\rho})-\log L(\widehat{\widehat{
		\lambda }}(\rho_{o}),\rho_{o})\right]\nonumber\\
&=-2\left[ x\log (\frac{x}{m}) - x +y\log(\frac{y}{n})
-y -x\log(\widehat{\widehat{
		\lambda}}(\rho_{o}))+m\widehat{\widehat{
		\lambda}}(\rho_{o})-\right.\nonumber\\
	& \left.{} y\log(\widehat{\widehat{
			\lambda}}(\rho_{o}) +\rho_{o})+n(\widehat{\widehat{
		\lambda}}(\rho_{o}) +\rho_{o})\right]
	\end{split}
\end{equation}
where $\widehat{\lambda},\widehat{\rho}$ are the maximum likelihood estimators of the parameters. The mle of cumulative excess deaths for $n$ after the emergency are:
\begin{equation}
n(\frac{y}{n}-\frac{x}{m})\label{eq:proflikexc}
\end{equation}
A well-known theorem by Wilks asserts that under some regularity conditions,\cite{pawitan01}  the profile likelihood 
ratio has a chi-square distribution with $1$ degree of freedom. A $(1-\alpha) \times 100$\%
confidence interval for $\rho$ can be found by moving from the top of the curve to the left and to the right until the function value has dropped by the $1-\alpha$ percentile of a chi-square
distribution with $1$ degree of freedom.\cite{rolkelopez01, rolkeetal05} For cumulative excess deaths, conservative simultaneous confidence intervals can be constructed using Bonferroni's method where each individual confidence coefficient is $1-\alpha/n$.

This model is useful in situations where there is limited data (e.g. only a few days), and the period of analysis has a constant background rate. If the emergency causes a large displacement of the population, or it covers a period of time where mortality rate changes, the proposed model could underestimate or overestimate excess deaths.

\subsection{Model 2: Log Linear Model}
Let $D_{j,t}=$ number of deaths at predictor combination $j$, time index $t$, $N_{j,t}=$ population size, and $\bolds{Z}_{j,t}$ a vector of predictors. Then, 
\begin{eqnarray}
\log(\mu_{j,t}) = \log(N_{j,t})+\beta_{o}+\bolds{Z}^{'}_{j,t} \bolds{\beta}\nonumber
\end{eqnarray}
where $\mu_{j,t}=E(D_t|\bolds{Z}^{'}_{j,t},N_{j,t})$. The natural logarithm of $N_{j,t}$ is an offset variable. Possible predictors include cause of death, age, gender, temporal trend, and seasonal variation. Santos and colleagues used a socioeconomic development index predictor, which captures the underlying strength of municipal level structural and institutional capacities, to estimate excess deaths due to Hurricane Maria.\cite{santosetal18} However, if the emergency event causes substantial damage to a location's infrastructure, it may be difficult to gather such data. Moreover, their method used monthly data although in general, excess deaths estimation is much more useful at a daily scale, since results could guide emergency management decisions.

Alternatively, let $D_{t}=$ number of deaths at time index $t$, $N_{t}=$ population size. For $l=0,...,L$, we use $p_{l, t}$ as an indicator of time period $l$ time $t$ falls in. These indicator variables permit us to estimate excess deaths; but where we allow the `excess death effect' to change in time. Specifically, $l=0$ represents the pre-emergency period; $l=1$, a period right after the emergency, and so forth.  Moreover, let $doy_t=$ day of year, and $year_t=$ a normalized year variable. Assuming $D_t$ follows a Poisson distribution, we propose the Generalized Additive Model (GAM), or more precisely, a semiparametric model \cite{ruppertetal03}; 
\begin{eqnarray}
\log(\mu_{l, t}) = \log(N_{l, t})+\beta_{o}+\bolds{p}^{'}_{t} \bolds{\beta} + f_{1}(doy_t)+f_{2}(year_t)\label{eq:model2}
\end{eqnarray}
where $\mu_{l, t}=E(D_t|t,\bolds{p}^{'}_{t},N_{l, t}, doy_t, year_t)$. The natural logarithm of $N_{l, t}$ is an offset variable; while $\bolds{p}^{'}_{t}=(p_{1,t}, \ldots, p_{L,t}),$ and $\bolds{\beta}=(\beta_1, \ldots, \beta_L)^{'}$ denote model coefficients for the $L$ post-Maria time periods. Meanwhile, $f_1$ is a smooth function of $doy$, which accounts for within year variation, while $f_2$ accounts for potential changes in demographics. For example, many populations are aging.\cite{bengtson18} $f_{1}$ is fit using a penalized cyclic cubic regression spline, and $f_2$ is fit using a penalized regression thin plate spline.\cite{wood17} Model coefficients are estimated by a penalized likelihood maximization approach, where the smoothing penalty parameters are determined by restricted maximum likelihood.\cite{wood11,reissandogden09} The model is fit using version 1.8-12 of the \textit{mgcv} package in \textsf{R}.\cite{wood17,rcran16} If residuals of the fitted model (\ref{eq:model2}) present remaining temporal dependence, a Generalized Additive Mixed Model will be considered.\cite{chienetal12,zwacketal11} 

The fit of model (\ref{eq:model2}) can be used to estimate excess deaths during time period $l$ that time index $t$ falls in, through the difference between the estimated model with $p_{l, t}=1$, versus the estimated model with $p_{l, t}=0$. For $l \ge 1$, let $\hat{\mu}_{l, t} = \widehat{E}(D_t|t,p_{l, t}=1,N_{l, t}, doy_t, year_t)$, $\hat{\psi}_{l, t} = \widehat{E}(D_t|t,p_{l, t}=0,N_{t}^{*}, doy_t, year_t)$, and $\hat{b}_{o}, \hat{b}_{l}$ estimate $\beta_{o}, \beta_{l}$ respectively. Also,  observe that $\hat{\psi}_{l, t}$ uses $N_t^{*}$, a population size unaltered by large post-emergency migration. Then,

\begin{eqnarray}
\hat{\mu}_{l, t} - \hat{\psi}_{l, t}&=&  \exp(\log(N_{l, t})+\hat{b}_{o}+ \hat{b}_{l} + \widehat{f}_1(doy_t)+\widehat{f}_2(year_t))\nonumber\\
&& -\exp(\log(N_t^{*}) +\hat{b}_{o}+ \widehat{f}_1(doy_t)+\widehat{f}_2(year_t))\nonumber\\
&=& e^{\hat{b}_{o}+\widehat{f}_1(doy_t)+\widehat{f}_2(year_t)}(e^{\log(N_{l, t})+\hat{b}_{l}}-e^{\log(N_t^{*})})\label{eq:excessd}
\end{eqnarray}
When $\bolds{p}^{'}_{t}=\bolds{0}$, then $\hat{\mu}_{l, t} - \hat{\psi}_{l, t}=0$. Equation (\ref{eq:excessd}) is the maximum likelihood estimator for expected excess deaths at $t$.\cite{casellaandberger02}

To estimate cumulative excess deaths for time period $l$; using (\ref{eq:excessd}),

\begin{eqnarray}
\sum_{t=q}^{r}(\hat{\mu}_{l, t} - \hat{\psi}_{l, t})\label{eq:cumdeaths}
\end{eqnarray} 
for any time period starting at index $q$ and ending at $r$.

\subsection{Model 2 Confidence Interval Through Posterior Simulation}
 Confidence intervals are commonly constructed using asymptotic approximations,\cite{ruppertetal03} bootstrapping,\cite{shabuzandgarthwaite18} and Bayesian methods.\cite{marraandwood12}  Our fitted penalized regression spline model does not lend itself to perform bootstrapping due to the smoothing penalty; but approximate simulations from the Bayesian posterior density of any function of coefficients are possible.\cite[p.~300]{wood17} In a nutshell, linear predictors are combined with smoothing bases and their penalties into one model matrix; while fixed effect coefficients and individual smooth term coefficients are stacked into vector $\bolds{\gamma}$. Then, assuming an improper prior distribution for $\bolds{\gamma}$, Wood shows that the approximate posterior distribution of $\bolds{\gamma}$ is\cite{wood17};

\begin{eqnarray}
\bolds{\gamma} \sim N(\hat{\bolds{\gamma}}, V_{\gamma})\label{eq:posterior}
\end{eqnarray}
where $\hat{\bolds{\gamma}}$ are the fitted coefficients and $V_{\gamma}$ is the covariance matrix of the coefficients. Strictly speaking, the posterior distribution above is conditional on the data, and a fixed smoothing parameter. Furthermore, intervals obtained from the posterior are Bayesian credible intervals, but they are commonly referred to as confidence intervals. \textsf{R} code is made available as supplementary material.


\subsection{Data}
After researchers and journalists were unable to obtain death certificate data from the Puerto Rico Vital Statistics System, in June 2018 a judge ordered the agency to make the data available to the public. Daily death certificate counts from January 1, 2015 until May 2018 were released. Causes of death and demographic data were not publicly released. The data set is dated until December 31, 2018, but is padded with zeros from May 31st onwards. Moreover, the most recent death certificate counts are likely preliminary; as evidenced by some May counts being exceedingly below average. Thus, we arbitrarily determine to only use death certificate counts from January 1, 2015 until February 28th, 2018.

Population data comes from the 2016 U.S. Census Vintage annual population estimates.\cite{census17} However, the life threatening conditions in Puerto Rico after Hurricane Maria's landfall convinced many residents to leave the island. For Model 2, we consider U.S. Bureau of Transportation Statistics data on monthly net movement of air passengers to adjust population estimates (Table \ref{tab:pass}).
Usually, net movement is a biased proxy of resident migration; mainly because it includes the seasonal movements of visitors,\cite{velazquezetal18} and tourism has been one of the few growing sectors in the Puerto Rican economy.\cite{rivera16} However, as Table \ref{tab:nrhotel} shows, a dramatic drop in visitors from the United States occurred the months following Maria's landfall (the U.S. dominates the local tourism market share). Therefore, net air passenger movement appears to be a reasonable proxy of migration for this period. According to net movement data, there were 145,623 more air passengers leaving the island than arriving at Puerto Rico from September 2017 to January 2018 (Table \ref{tab:pass}). In recent years, the Puerto Rican population has been decreasing due to economic woes, but not at the scale of the months post-Maria. 

\begin{table}[H]
	\footnotesize
	\centering
	\scalebox{0.7}{
		\begin{tabular}{cccc}
			\hline
			\textbf{Month} &  \textbf{Passengers Leaving} & \textbf{Passengers Arriving} & \textbf{Net Movement}\\[5pt] 
			Sep-2017 &  194571 &    149848 &    44723 \\[5pt] 
			Oct-2017 &  258662 &	159465 &	99197\\[5pt] 
			Nov-2017 &  265606 &	215356 &	50250\\[5pt]
			Dec-2017 &  354865 &	332710 &	22155\\[5pt]
			Jan-2018 &  289231 &	359921 &   -70690\\[5pt]\hline
			&&\textbf{Total}& 145,623\\\hline            
		\end{tabular}
	}
	\caption{Air passenger movement from and to Puerto Rico. Source: U.S. Bureau of Transportation Statistics. Accessed July 11, 2018 from \url{https://indicadores.pr}}
	\label{tab:pass}
\end{table}

\begin{table}[H]
	\footnotesize
	\centering
	\scalebox{0.7}{
		\begin{tabular}{cc}
			\hline
			\textbf{Month} & \textbf{Non-resident Registration Change (\%)}\\[5pt] 
			Sep-2017  & -25.0\\[5pt] 
			Oct-2017 & -75.2\\[5pt] 
			Nov-2017 & -76.2\\[5pt]
			Dec-2017 & -80.8\\[5pt]
			Jan-2018 & -61.6\\[5pt]\hline           
		\end{tabular}
	}
	\caption{Percentage change in non-resident hotel registrations originating from the U.S. relative to the same period in the previous year. Source: Puerto Rico Tourism Company: \url{https://www.prtourism.com/dnn/Statistics_old02162018}.}
	\label{tab:nrhotel}
\end{table}

No air passenger data was available for February; we assumed the population for this month was the same as January. Thus, for Model 2, daily population estimates are determined as follows:
\begin{itemize}
	\item First, get the end of September 2017 population estimate: the September 2017 net movement is subtracted from the 2016 U.S. Census Vintage population estimate, 3337177, for 2017.
	\item For the end of October 2017 population estimate, the October 2017 net movement is subtracted from the September 2017 population estimate. 
	\item End of November, December, and January population estimates are obtained analogously to the step above.
	\item These population estimates are combined with U.S. Census Vintage population estimates to interpolate daily populations. 
\end{itemize} 

\section{Results}
\begin{figure}[H]
	\begin{center}

		\includegraphics[width=3.8in]{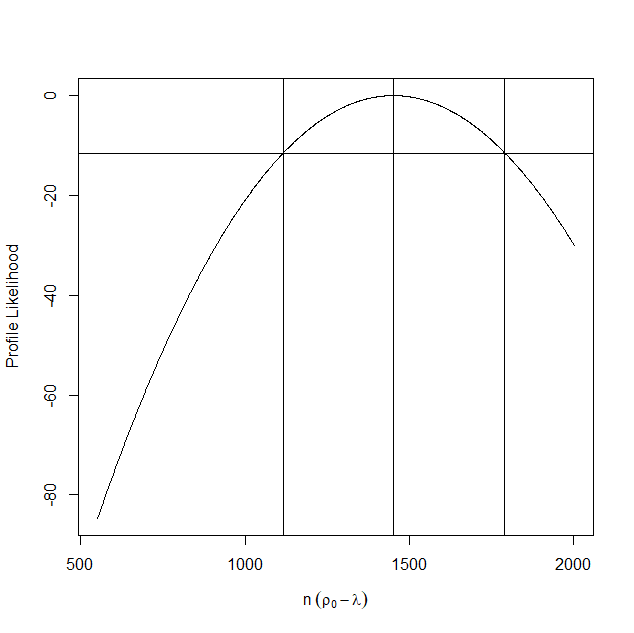}
	\end{center}
	\caption{Profile Likelihood Hurricane Maria cumulative excess deaths estimate covering the period from September 20, until November 30, 2017. The horizontal lines in each side of the peak of the curve represent 95\% CI bounds.} \label{fig:proflikNov30}
\end{figure}

We implement Model 1 in order to quickly assess excess deaths a short period after Hurricane Maria made landfall in Puerto Rico. Specifically, exploratory analysis suggested that before Hurricane Maria, there was little within year variation among most monthly death rates with noticeably higher death rates in the months of December and January. Moreover, year to year raw death rates may be increasing. We performed analysis of variance at 5\% significance, to determine if mean number of deaths for the months of May to August 2017 were equal.
  With a p-value of $0.40$, the null was not rejected. Residual analysis did not indicate any concern with ANOVA assumptions. Therefore, 2017 daily deaths from May 1, until September 19 are used to estimate the background death rate before Hurricane Maria. Cumulative excess deaths estimates (Figure \ref{fig:proflikNov30}) using Model 1 will be presented later on.

\begin{figure}[H]
	\begin{center}

		\includegraphics[width=3.8in]{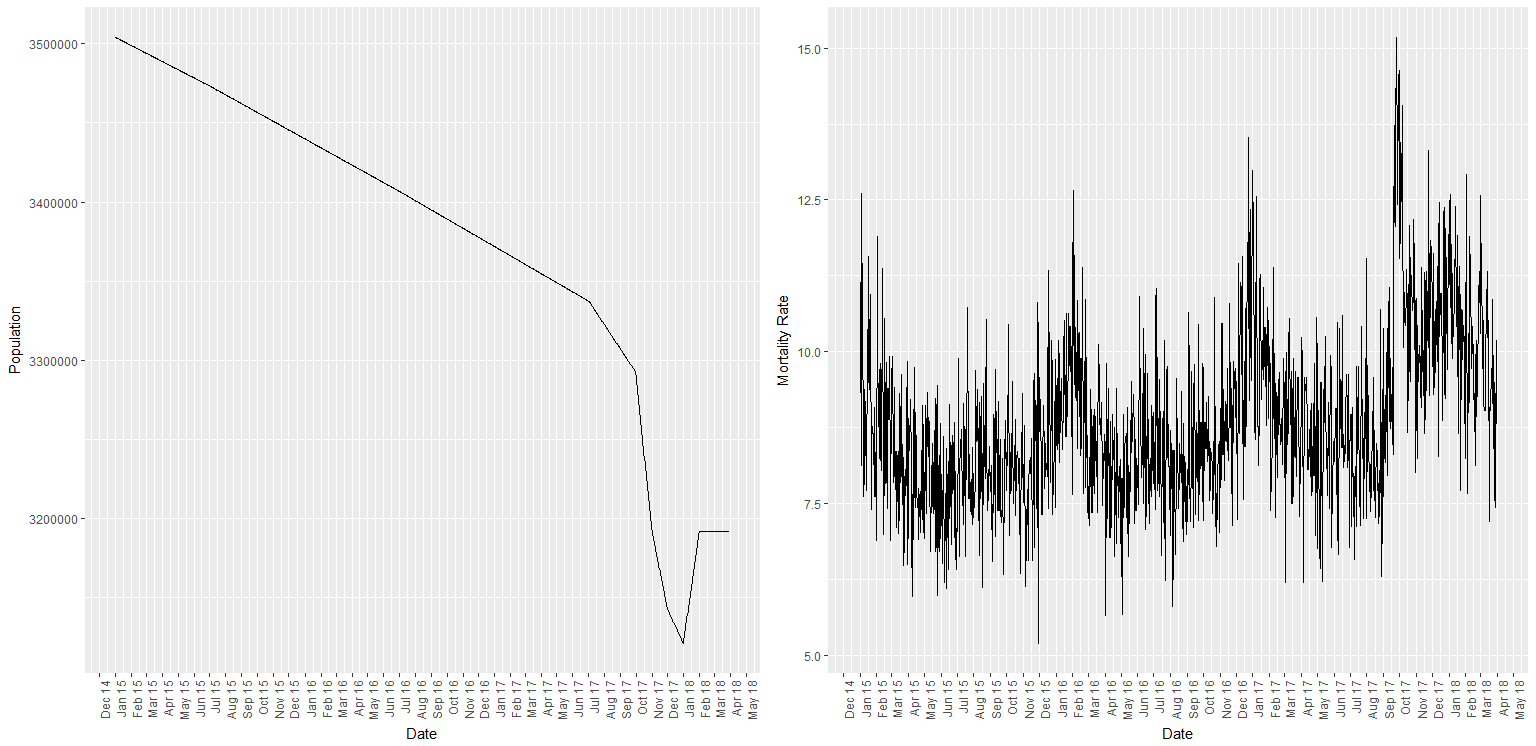}
	\end{center}
	\caption{Left panel displays daily population estimates from January 1, 2015 until February 28, 2018. Right panel displays daily mortality rates for the same time period.} \label{fig:popandMortalityrate}
\end{figure}

For Model 2, our daily population estimates account for changes in population (Figure \ref{fig:popandMortalityrate}, left panel). Moreover, end of month population estimates for September, October, November, December, and January were: 1.34, 4.31, 5.82, 6.48, and 4.36 percent smaller than the Census Vintage population estimate for mid-2017. A quickly decreasing population may also affect daily death counts. Thus, not accounting for the decrease in population could hinder the estimation of the Hurricane Maria death toll in uncertain ways. Daily death certificate data was combined with population estimates to produce daily mortality rates; 
\begin{eqnarray}
R_{t} = \frac{D_t}{N_t}\times 1000\times 365\nonumber
\end{eqnarray}
From Figure \ref{fig:popandMortalityrate}, right panel, it is seen that the most pronounced temporal dependence in mortality rates was generally higher values in January and December; and a dramatic jump in mortality rates in September 2017, which remained abnormally high for a portion of the rest of the year. It is unclear from the mortality rate chart if mortality rates were increasing across years; yet $f_{2}$ in (\ref{eq:model2}) can account for this type of temporal fluctuation. 

Our version of model (\ref{eq:model2}) included indicators for the days after Hurricane Maria in each month from September 2017 to February 2018. After exploratory analysis, each regression spline basis dimension was set at 32; and $year$ was modeled linearly. All inference was conducted at 5\% significance. The sensitivity of our results to the population adjustment from air passenger net movement was assessed by comparing results without implementing this adjustment. Table \ref{tab:pvalues} presents Wald test p-values. The findings are that the post-Maria days in 2018 months are not significant when net movement adjustment is applied; without population adjustment, no December 2017 effect was found. But Wald tests are based on the asymptotic normal assumption and dependent on model parameterization.\cite{wood17} 
Thus, we proceeded by performing a generalized likelihood ratio test,\cite{suarezetal17} to compare a model with September, October, November and December; versus a model with only September, October, and November.  Table \ref{tab:lrt} corroborates that with net movement adjustment, the simpler model should be rejected; without the net movement adjustment, the simpler model was preferred.


\begin{table}[H]
	\footnotesize
	\centering
	\scalebox{0.7}{
		\begin{tabular}{ccc}
			\hline
			& \textbf{Adjusted} & \textbf{No Adjustment}\\[5pt]\hline 
			\textbf{Fixed Effect} &  \textbf{p-value} & \textbf{p-value}\\[5pt] 
			Sep (20 to 30th) &  0 &0\\[5pt] 
			Oct-2017 &  0 &0\\[5pt] 
			Nov-2017 &  0 &0\\[5pt]
			Dec-2017 &  0.004 &0.512\\[5pt]
			Jan-2018 &  0.446 &0.293\\[5pt]
			Feb-2018 &  0.369 &0.769\\[5pt]
			\hline           
		\end{tabular}
	}
	\caption{Wald test results for statistical significance of post-Maria periods. First p-values are when population is adjusted with net movement. Second p-values are when population is not adjusted.}
	\label{tab:pvalues}
\end{table}

\begin{table}[H]
	\footnotesize
	\centering
	\scalebox{0.7}{
		\begin{tabular}{cccc}
			\hline
			&& \textbf{Adjusted} & \textbf{No Adjustment}\\[5pt]\hline
			\textbf{Model} &  \textbf{Variables} &\textbf{p-value} &\textbf{p-value}\\[5pt] 
			1 & September, October, November, December & 0.008 & 0.522\\[5pt]
			2 & September, October, November & \\[5pt]
			\hline           
		\end{tabular}
	}
	\caption{Generalized Likelihood Ratio test results when population is adjusted for net movement (first column), and when it is not (second). Simpler model is in the null.}
	\label{tab:lrt}
\end{table}

We should remark that non-significance does not mean that people did not die due to Hurricane Maria past December; but that if deaths did occur, they did not occur at a high enough rate to be detected through our model. For example, there is strong evidence that two deaths that occurred in January and February 2018 may have been associated to the storm yet were not part of the official death toll.\cite{sutteretal18} Henceforth, only results from the net movement adjusted data are presented.

\begin{figure}[H]
	\begin{center}

		\includegraphics[width=3.8in]{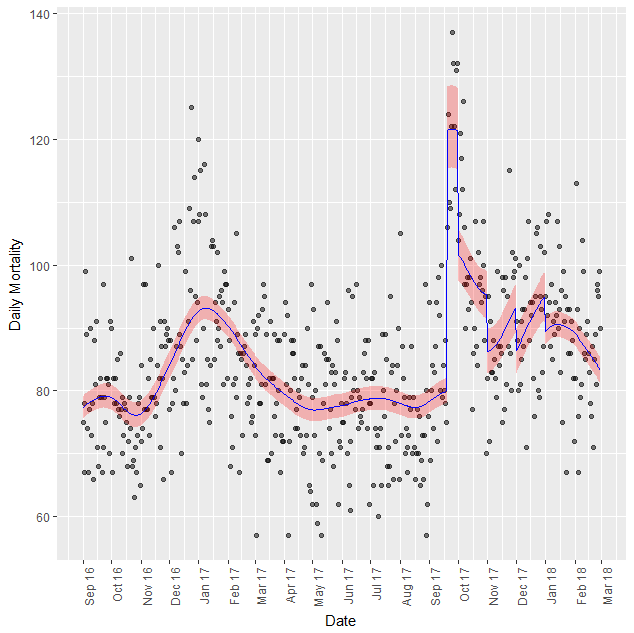}
	\end{center}
	\caption{Daily mortality rate from September 1, 2016 until February 28, 2018. The corresponding model fit (blue line), and simultaneous confidence band for the curve.} \label{fig:fitperformance}
\end{figure}

Table \ref{tab:coef} shows the coefficient estimates for each statistically significant post-Maria period. Specifically, from September 20th to September 30th, it was estimated that residents had 1.517 times the risk of dying compared to the pre-Maria period. Moreover, October, November, December increased the pre-Maria mortality rate by 27.2\%, 15.0\% and 6.4\% respectively. 
\begin{table}[H]
	\footnotesize
	\centering
	\scalebox{0.7}{
		\begin{tabular}{ccc}
			\hline
			\textbf{Fixed Effect} &  \textbf{Estimated Coefficient (S.E.)}&\textbf{Multiplicative Effect}\\[5pt] 
			Sep (20 to 30th) &  0.418 (0.030) & 1.517\\[5pt] 
			Oct-2017 &  0.241 (0.022) & 1.272\\[5pt] 
			Nov-2017 &  0.140 (0.023) &1.150\\[5pt]
			Dec-2017 &  0.062 (0.022) &1.064\\[5pt]
			\hline           
		\end{tabular}
	}
	\caption{Estimated model coefficients (standard errors are in parenthesis). Estimated mortality multiplicative effects of post-Maria periods.}
	\label{tab:coef}
\end{table}
Figure \ref{fig:fitperformance} presents daily mortality, model fits, and simultaneous confidence bands. Model diagnostics (not shown) did not indicate overdispersion, nor temporal dependence remaining in the residuals.

\subsection{Cumulative Excess Deaths}
Using (\ref{eq:proflikexc}), (\ref{eq:cumdeaths}) and (\ref{eq:posterior}), Table \ref{tab:cdeath} shows point-wise confidence intervals for cumulative excess deaths at the end of each statistically significant post-Maria month. Thus, our Model 2 indicates with 95\% confidence that in the first 14 weeks after Hurricane Maria, between 1,069 and 1,568 people died directly, or indirectly due to the storm. Qualitatively, Model 1 and Model 2 confidence intervals overlap. However, Model 2 results in generally narrower confidence intervals.

\begin{figure}[H]
	\begin{center}

		\includegraphics[width=3.8in]{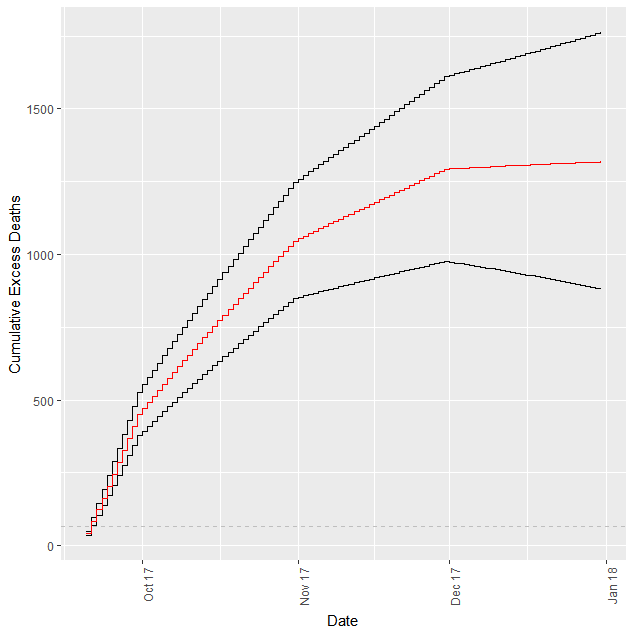}
	\end{center}
	\caption{Model 2 cumulative excess deaths estimates from September 20th, 2017, until February 28, 2018. Step function in red is the estimated expected excess deaths. Step functions in black are lower and upper 95\% confidence bounds.} \label{fig:cumulative_excess_deaths}
\end{figure}

It is best to account for uncertainty in the estimated excess curve through a simultaneous confidence band (Figure \ref{fig:cumulative_excess_deaths}). On average, across the post-Maria period presented, according to Model 2 we are 95\% confident the band includes the true cumulative excess death function.
\begin{table}[H]
	\footnotesize
	\centering
	\scalebox{0.7}{
		\begin{tabular}{ccccc}
			\hline
			\textbf{Period} & \textbf{Model 1 Estimate}&\textbf{95\% C.I.}& \textbf{Model 2 Estimate}&\textbf{95\% C.I.}\\[5pt] 
			Sep 20 - Sep 30 & 482 &(358, 613)&449 & (377, 527)\\[5pt] 
			Sep 20 - Oct 31 & 1112 & (867, 1363)&1046 & (893, 1197)\\[5pt] 
			Sep 20 - Nov 30 & 1453 & (1116, 1791) &1293 & (1086, 1495)\\[5pt]
			Sep 20 - Dec 31 & - & - & 1318 & (1069, 1568)\\[5pt]
			\hline           
		\end{tabular}
	}
	\caption{Estimates of cumulative excess deaths from both models, and their respective 95\% confidence intervals. }
	\label{tab:cdeath}
\end{table}

\section{Discussion}
In this paper, models that estimate excess deaths were proposed. Model 1 performs a simple before and after analysis using the profile likelihood method. Model 2 allows statistical inference on the duration of the effect, and can incorporate population displacement into excess deaths estimation.
Furthermore, Model 2 accounts for uncertainty on the excess death curve using a simultaneous confidence band. The statistical procedures have few modeling assumptions.  They can be used to estimate excess deaths due to an emergency when it is difficult to follow CDC guidelines, or to assess guideline based estimates. Changes in disease incidence can be modeled with our method as well. Moreover, our Model 2 used air passenger data to adjust mortality rates due to changes in the population size. \textsf{R} code is made available as supplementary material. We also explored using an adaptive smoothing penalty in Model 2,\cite{hastieetal09} but this did not improve results. When inferring on the duration of the post-emergency effect, an implicit assumption in our method is that death rates will return to pre-emergency values.  We can easily imagine emergency events that lead to a new death rate behavior; hindering this type of inference. Moreover, our Poisson GAM does not account for uncertainty in the smoothing parameter, nor population estimates. A fully Bayesian model could overcome these limitations, but face considerable computational challenges.\cite{gelmanetal13, krivobokovaetal10,fahrmeiretal04}

We have analyzed the number of deaths in Puerto Rico before and after Hurricane Maria. Our Model 2 indicates that with 95\% confidence, the total death toll was between 1,069, and 1,568. Moreover, the aftermath from Hurricane Maria led to decreasing excess deaths until sometime in December. Deaths outside of Puerto Rico that would have not occurred had Maria not made landfall are not assessed in this study. Yet, it should be noted that only a small percentage of out of state deaths after Hurricane Katrina were classified as Katrina-related.\cite{brunkardetal08} The difficulty in treating noncommunicable diseases after Hurricane Maria may explain the large death toll. With individual death certificate data that includes cause of death, age, gender, and location of residence, our model can be modified to track the profile of common indirect causes of death after natural disaster such as diabetes, leptospirosis, and other conditions. Interaction terms between the causes of death and the indicator variables, could also be incorporated. Unfortunately, individual registries were made available only recently. Future research will assess mortality due to noncommunicable diseases.

Santos and Howard compared historical monthly deaths with September through December 2017 deaths to estimate the Hurricane Maria death toll in that time frame.\cite{santosandhoward18} Their method did not use any population data and assumed that death counts did not vary across years. Rivera and Rolke compared deaths for the first 19 days of September 2017 with deaths from the first 6 weeks after landfall to estimate the Hurricane Maria death toll in that period.\cite{riveraandrolke18} Their method made no adjustments for population loss, and relied on preliminary September-October death certificate data.  Kishore and colleagues conducted household surveys and then extrapolated from sample estimates to infer on the number of people who died through December 31, 2017.\cite{kishoreetal18} Perhaps because their sample size was relatively low for the task, their confidence interval had a very large margin of error.  

Santos and colleagues used individual-level data including age, gender, temporal trend, seasonal variation, and a socioeconomic development index (SEI) predictor, which captures the underlying strength of municipal level structural and institutional capacities, to estimated excess deaths.\cite{santosetal18} However, this data was exclusively made available to the researchers almost six months after Hurricane Maria struck. Moreover, the SEI tertiles were based on 2013 data, and an error led to an overestimation by over 300 deaths. Although the authors state that the error only affected their January and February 2018 excess estimates, the December 2017 displacement numbers they present are twice as large as ours. It should also be noted that all of these other studies indicate that the Hurricane Maria death toll was far above 64. Yet, none can draw inference on whether the post-Maria death toll effect dissipated; as the method proposed in this paper can.

Direct forensic data are desirable to understand the causes of deaths from a disaster, but such data are subject to certain weaknesses, such as ambiguity in determining whether the death is directly related to the disaster.  Ideally, forensic assessments can be combined with indirect estimates of the numbers of excess deaths, as provided in our paper, to gain a more complete understanding of the nature and extent of the casualties.

\section*{Acknowledgments}
\thispagestyle{empty}
The authors would like to thank the anonymous referees and Associate Editor for their helpful recommendations.

\clearpage

\appendix
%
%
%
The log-likelihood of the pre-emergency and post-emergency total deaths is:
\begin{equation*}
\begin{split}
\log L&=K+x\log (\lambda )-m\lambda +y\log (\lambda+\rho  )- n(\lambda+\rho  ) \\
&\frac{d\log L}{d\rho }=\frac{x}{\lambda + \rho}-n=0 \text{ , } \widehat{\rho }=\frac{y}{n}-\lambda \\
&\frac{d\log L}{d\lambda }=\frac{x}{\lambda }-m+\frac{y}{\lambda+\rho }-n =0 \text{ , } \widehat{\lambda }=x/m
\end{split}
\end{equation*}

where $K$ is a constant. For the likelihood ratio test:

\begin{equation*}
\begin{split}
\frac{d\log L}{d\lambda }&=\frac{x}{\lambda }-m+\frac{y}{\lambda+\rho_{o} }-n =0 \\ 
\widehat{\widehat{
		\lambda }} (\rho_{o})&=\left( x+y -(m+n) \rho_{o} \pm \sqrt{(x+y-(m+n)\rho_{o})^{2}+4(m+n)x\rho_{o} }\right) /\left( 2(m+n )\right)
\end{split}
\end{equation*}

\bibliographystyle{asa}
\bibliography{bibtexreferences}

\end{document}